\documentstyle[12pt,epsf]{article}
\title{The Sum Rules for Structure
Functions of Polarized $e(\mu)N$ Scattering: Theory vs. Experiment}
\author{B.L.Ioffe\\
Institute of Theoretical and Experimental Physics\\
117259, Moscow, Russia}
\date{}
\begin{document}
\maketitle

\newcommand{\be}{\begin{equation}}
\newcommand{\ee}{\end{equation}}

\def\la{\mathrel{\mathpalette\fun <}}
\def\ga{\mathrel{\mathpalette\fun >}}
\def\fun#1#2{\lower3.6pt\vbox{\baselineskip0pt\lineskip.9pt
\ialign{$\mathsurround=0pt#1\hfil##\hfil$\crcr#2\crcr\sim\crcr}}}


\vspace{1cm}

\begin{abstract}

The determination
of twist-4 corrections to the structure functions of polarized
$e(\mu)N$  scattering by QCD sum
rules is reviewed and critically analyzed. It is found that in the case of
the Bjorken sum rule the twist-4 correction is small at \\
$Q^2 > 5~GeV^2$.
However, the accuracy of the today experimental data is
insufficient to reliably determine $\alpha_s$ from the Bjorken sum rule. For
the singlet sum rule -- $p+n$ -- the QCD sum rule gives only the order of
magnitude of twist-4 correction. At low and intermediate $Q^2$ the model is
presented which realizes a smooth connection of the Gerasimov-Drell-Hearn
sum rules at $Q^2 = 0$ with the sum rules for $\Gamma_{p,n}(Q^2)$ at high
$Q^2$. The model is in a good agreement with the experiment.
\end{abstract}

\newpage
{\large \bf 1.~ Introduction}

\vspace{2mm}
In the last few years there is a strong interest to the problem of nucleon
spin structure: how nucleon spin is distributed among its constituents - quarks
and gluons. New experimental data continuously appear and precision
increases (for the recent data see [1], [2]). One of the most
important item of the information comes from the measurements of the first
moment of the spin-dependent nucleon structure functions $g_1(x)$ which
determine the parts of nucleon spin carried by $u,~ d$ and $s$ quarks and
gluons.  The accuracy of the data is now of a sort that the account of
twist-4 terms is of importance when comparing the data with the Bjorken
and Ellis-Jaffe sum rules at high $Q^2$. On the other side,  at low and
intermediate $Q^2$ a smooth connection of the sum rules for the first
moments of $g_1(x,Q^2)$ with the Gerasimov-Drell-Hearn (GDH) sum
rules [3,4] is theoretically expected. This connection can be realized
through nonperturbative $Q^2$-dependence only. In my talk I discuss such
nonperturbative $Q^2$-dependence of the sum rules (see also [5]).

Below I
will consider only the first moment of the structure function $g_1(x,Q^2)$

\be
\Gamma_{p,n}(Q^2) = \int \limits^{1}_{0}~ dx~g_{1; p,n}(x, Q^2)
\ee

The presentation of the material is divided into two parts. The
first part deals with the case of high $Q^2$. I discuss the determination
of twist-4 contributions to $\Gamma_{p,n}$ by QCD sum rules and with
the account of twist-4 corrections and of the uncertainties in their values
compare the theory with the experiment. In the second part the case of low and
intermediate $Q^2 \la 1 GeV^2$ is considered in the framework of the model,
which realizes the smooth connection of GDH sum rule at $Q^2 = 0$ with the
asymptotic form of $\Gamma_{p,n}(Q^2)$ at high $Q^2$.

\vspace{3mm}
{\large \bf 2. ~High $Q^2$}

\vspace{2mm}
At high $Q^2$ with the account of twist-4 contributions $\Gamma_{p,n}(Q^2)$
have the form

\be
\Gamma_{p,n}(Q^2) = \Gamma^{as}_{p,n}(Q^2) + \Gamma^{tw 4}_{p,n}(Q^2)
\ee
$$
\Gamma^{as}_{p,n}(Q^2) = \frac{1}{12} \Biggl \{ [1 - a - 3.58 a^2 - 20.2 a^3
- c a^4 ] [\pm g_A + \frac{1}{3}a_8]
$$

\be
+ \frac{4}{3} [1 - \frac{1}{3} a - 0.55 a^2 - 4.45 a^3] \Sigma \Biggr \} -
\frac{N_f}{18 \pi} \alpha_s(Q^2) \Delta g(Q^2)
\ee

\be
\Gamma^{tw4}_{p,n}(Q^2) = \frac{b_{p,n}}{Q^2}
\ee
In eq.(3) $a = \alpha_s(Q^2)/\pi, g_A$ is the $\beta$-decay axial coupling
constant, $g_A = 1.260 \pm 0.002$ [6]

\be
g_A = \Delta u - \Delta d ~~~~ a_8 = \Delta u + \Delta d - 2 \Delta s ~~~~
\Sigma = \Delta u + \Delta d + \Delta s.
\ee
$\Delta u, \Delta d, \Delta s, \Delta g$ are parts of the nucleon spin
projections carried by $u, d, s$ quarks and gluons:

\be
\Delta q = \int \limits^{1}_{0} \Biggl [q_+(x) - q_-(x) \Biggr ]
dx
\ee
where $q_+(x), q_-(x)$ are quark distributions with spin projection
parallel (antiparallel) to nucleon spin and a similar definition takes place
for $\Delta g$. The coefficients of perturbative series were calculated in
[7-10], the numerical values in (3) correspond to the number of flavours
$N_f = 3$, the coefficient $c$ was estimated in [11], $c \approx 130$. In
the $\overline{MS}$ renormalization scheme chosen in [7-10] $a_8$ and $\Sigma$ are
$Q^2$-independent. In the assumption of the exact $SU(3)$ flavour symmetry
of the octet axial current matrix elements over baryon octet states $a_8 =
3F - D = 0.59 \pm 0.02$ [12].

Strictly speaking, in (3) the separation of terms proportional to $\Sigma$
and $\Delta g$ is arbitrary, since the operator product expansion (OPE) has
only one singlet in flavour twist-2 operator for the first moment of the
polarized structure function -- the operator of singlet axial current
$j^{(0)}_{\mu 5}(x) = \sum \limits_{q}~ \bar{q}_i(x) \gamma_{\mu} \gamma_5 q, ~~ q =
u,d,s$. The separation of terms proportional to $\Sigma$ and $\Delta g$ is
outside the framework of OPE and depends on the infrared cut-off. The
expression used in (3) is based on the physical assumption that the
virtualities $p^2$ of gluons in the nucleon are much larger than light quark
mass squares, $\vert p^2 \vert \gg m^2_q$ [13] and that the infrared cut-off is chosen
in a way providing the standard form of axial anomaly [14].

Twist-4 corrections to $\Gamma_{p,n}$
were calculated by Balitsky, Braun and
Koleshichenko (BBK) [15]  using the QCD sum rule method.

BBK calculations were critically analyzed in [5], where it was shown
that there are few possible uncertainties in these calculations: 1)
the main contribution to QCD sum rules comes from the last accounted
term in OPE -- the operator of dimension 8; 2) there is a large
background term and a much stronger influence of the continuum
threshold comparing with usual QCD sum rules; 3) in the singlet case, when
determining the induced by external field vacuum condensates, the corresponding
sum rule was saturated by $\eta$-meson, what is wrong. The next order
term -- the contribution of the dimension 10 operator in the BBK sum
rules was estimated by Oganesian [16]. The account of the dimension-10 contribution
to the BBK sum rules and estimation of other uncertainties results in
(see [5]):

\be
b_{p-n} = -0.006 \pm 0.012~GeV^2
\ee
\be
b_{p+n} = -0.035 (\pm 100 \%)~GeV^2
\ee
As is seen from (7), in the nonsinglet case the twist-4 correction is
small ($\la 2\%$ at $Q^2 \ga 5 GeV^2$)
even with the account of the
error. In the singlet case the situation is much worse: the estimate (8)
may be considered only as correct by the order of magnitude.

I turn now to comparison of the theory with the recent experimental data.
In Table 1 the recent data obtained by SMC [1] and E~154(SLAC) [2]
groups are presented.

\newpage
\vspace{3mm}
\centerline{\large \bf Table ~ 1}

\vspace{2mm}
\begin{tabular}{|c|c|c|c|c|}\hline
&     $\Gamma_p$ & $\Gamma_n$ & $\Gamma_p - \Gamma_n$  & $\alpha_s(5
GeV^2)$\\ \hline
SMC & $0.132 \pm 0.017$ & $ -0.048 \pm0.022$ & $0.181
\pm 0.035$ & $0.270^{+ 0.16}_{-0.40}$\\
\hline
combined & $0.142 \pm 0.011$ & $-0.061 \pm 0.016$ & $0.202 \pm 0.022$
& $0.116^{+ 0.16}_{-0.44}$\\ \hline
E 154(SLAC) & $0.112 \pm 0.014$ & $-0.056 \pm 0.008$ & $0.168 \pm
0.012$ & 0.339 $^{+ 0.052}_{-0.063}$\\ \hline
EJ/Bj sum rules & $0.168 \pm 0.005$ & $-0.013 \pm 0.005$ & $0.181 \pm 0.002$ & 0.276\\ \hline
\end{tabular}

\vspace{3mm}
In the second line of Table 1 the results of the performed by SMC [1]
combined analysis of SMC [1], SLAC-E80/130 [17], EMC [18] and SLAC-E143
[19] data are given. The data presented in Table 1 refer to $Q^2 = 5
GeV^2$.  In all measurements each range of $x$ corresponds to each own
mean $Q^2$. Therefore, in order to obtain $g_1(x, Q^2)$ at fixed $Q^2$
refs. [1,2] use the following procedure. At some reference scale
$Q^2_0$ ($Q^2_0 = 1 GeV^2$ in [1] and $Q^2_0 = 0.34 GeV^2$ in [2]) quark
and gluon distribution were parametrized as functions of $x$. (The
number of the parameters was 12 in [1] and 8 in [2]). Then NLO
evolution equations were solved and the values of the parameters were
determined from the best fit at all data points. The numerical values
presented in Table 1 correspond to $\overline{MS}$ regularization
scheme, statistical, systematical, as well as theoretical errors arising
from uncertainty of $\alpha_s$ in the evolution equations, are added in
quadratures. In the last line of Table 1 the Ellis-Jaffe (EJ) and
Bjorken (Bj) sum rules prediction for $\Gamma_p, \Gamma_n$ and
$\Gamma_p -\Gamma_n$, correspondingly are given. The EJ sum rule
prediction was calculated according to (3), where $\Delta s = 0$ ,
i.e., $\Sigma = a_8 = 0.59$ was put and the last-gluonic term in (3)
was omitted. The twist-4 contribution was accounted in the Bj sum rule
and included into the error in the EJ sum rule. The $\alpha_s$ value in
the EJ and Bj sum rules calculation was chosen as $\alpha_s(5 GeV^2) =
0.276$, corresponding to $\alpha_s(M_z) = 0.117$ and
$\Lambda^{(3)}_{\overline{MS}} = 360 MeV$ (in two loops). As is clear
from Table 1, the data, especially for $\Gamma_n$, contradict the EJ sum
rule. In the last column, the values of $\alpha_s$ determined from the
Bj sum rule are given with the account of twist-4 corrections.

The experimental data on $\Gamma_p$ presented in Table 1 are not in a
good agreement. Particularly, the value of $\Gamma_p$ given by E154
Collaboration seems to be low: it does not agree with the old data
presented by SMC [20] ($\Gamma_p = 0.136 \pm 0.015$) and E143 [19]
($\Gamma_p = 0.127 \pm 0.011$). Even more strong discrepancy is seen in
the values of $\alpha_s$, determined from the Bj sum rules. The value
which follows from the combined analysis is unacceptably low: the
central point corresponds to $\Lambda^{(3)}_{\overline{MS}} = 15 MeV$!
On the other side, the value, determined from the E154 data seems to be
high, the corresponding $\alpha_s(M_z) = 0.126 \pm 0.009$. Therefore, I
come to a conclusion that at the present level of experimental accuracy
$\alpha_s$ cannot be reliably determined from the Bj sum rule in
polarized scattering.

Table 2 shows the values of $\Sigma$ -- the total nucleon spin
projection carried by $u, d$ and $s$-quarks found from $\Gamma_p$ and
$\Gamma_n$
presented in Table 1 using eq.(3). (It was put $g_A = 1.260, a_8 =
0.59$, the term, proportional to $\Delta g$ is included into $\Sigma$.).

\newpage

\centerline{\large \bf Table ~2: ~~The values of $\Sigma$}

\vspace{3mm}
\noindent
\vspace{2mm}
\begin{tabular}{|c|c|c|c|c|}\hline
& \multicolumn{2}{|c|}{From $\Gamma_p$}
& \multicolumn{2}{|c|}{From $\Gamma_n$}\\ \hline
& At $\alpha_s(5 GeV^2)=$ & At $\alpha_s( GeV^2)$ & At $\alpha_s(5 GeV^2)=$& At $\alpha_s(5 GeV^2)$\\
&$ =0.276$ & given in Table 1 & $ =
0.276$
 & given in Table 1\\ \hline
SMC & 0.256 & 0.254 & 0.264 & 0.266\\ \hline
Comb. & 0.350 & 0.250 & 0.145 & 0.225 \\ \hline
E154 & 0.070(0.14; 0.25) & 0.133(0.20; 0.30) & 0.19(0.25; 0.14) & 0.144
(0.205; 0.10)\\ \hline
\end{tabular}

\vspace{3mm}
In their fitting procedure [2] E154 Collaboration used the values $a_8
= 0.30$ and $g_A = 1.09$. The values of $\Sigma$ obtained from
$\Gamma_p$ and $\Gamma_n$ given by E154 at $a_8 = 0.30, g_A = 1.26$ and
$a_8 = 0.30, g_A = 1.09$ are presented in parenthesis. The value $a_8 =
0.30$ corresponds to a strong violation of SU(3) flavour symmetry and
is unplausible; $g_A = 1.09$ means a bad violation of isospin and is
unacceptable. As seen from Table 1, $\Sigma$ is seriously affected by
these assumptions. The values of $\Sigma$ found from $\Gamma_p$ and
$\Gamma_n$ using SMC and combined analysis data agree with each other
only,if one takes for $\alpha_s(5 GeV^2)$ the values given in\\ Table 1
($\alpha_s = 0.116$ for combined data), what is unacceptable. The
twist-4 corrections were not accounted in $\Sigma$ in Table 2: their
account, using eqs.(7), (8), results in increasing of $\Sigma$ by
0.04 if determined from $\Gamma_p$ and by 0.03 if determined from $\Gamma_n$.

To conclude, one may say, that the most probable value of $\Sigma$ is
$\Sigma \approx 0.3$ with an uncertain error. The contribution of
gluons may be estimated as $\Delta g (1 GeV^2) \approx 0.3$ (see [5],
[21], [22]). Then $\Delta g(5 GeV^2) \approx 0.6$ and the account of
gluonic term in eq.(3) results in increasing of $\Sigma$ by
0.06. At $\Sigma = 0.3$ we have $\Delta u = 0.83, \Delta d = -0.43,
\Delta s = -0.1$.

\vspace{3mm}
\centerline {\large \bf 3. ~Low and Intermediate $Q^2$}

\vspace{2mm}
The problem of a smooth connection of the Gerasimov-Drell-Hearn (GDH)
sum rules [3,4] which holds at $Q^2 = 0$, and the sum rules at high
$Q^2$ attracts attention in the last years [5, 23-25].

In order to connect the GDH sum rule with $\Gamma_{p,n}(Q^2)$ consider
the integrals [26]

\be
I_{p,n}(Q^2) =
\int\limits^{\infty}_{Q^2/2}~\frac{d\nu}{\nu}G_{_1;p,n}(\nu,Q^2)
\ee
Changing the integration variable $\nu$ to $x$, (9) can be
also identically written as

\be
I_{p,n}(Q^2) = \frac{2 m^2}{Q^2}~\int\limits^1_0~dx~ g_{1;p,n} (x,Q^2) =
\frac{2m^2}{Q^2}\Gamma_{p,n} (Q^2)
\ee
At $Q^2 = 0$ the GDH sum rule takes place

\be
I_p(0) = -\frac{1}{4}\kappa^2_p = - 0.8035;~~~I_n (0) =
-\frac{1}{4}\kappa^2_n = =0.9149; ~~~I_p(0) - I_n(0) = 0.1114
\ee
where $\kappa_p$ and $\kappa_n$ are proton and neutron anomalous
magnetic moments.

The schematic $Q^2$  dependence of $I_p(Q^2),~I_n(Q^2)$  and $I_p(Q^2) -
I_n(Q^2)$  is plotted in Fig.1. The case of $I_p(Q^2)$ is especially
interesting: $I_p(Q^2)$  is positive, small and decreasing at $Q^2 \ga
3GeV^2$  and negative and relatively large in absolute value at $Q^2=0$.
With $I_n(Q^2)$ the situation is similar. All this indicates  large
nonperturbative effects in $I(Q^2)$ at $Q^2 \la 1GeV^2$.

In [23]  the model was suggested, which describes $I(Q^2)$ (and
$\Gamma(Q^2)$) at low and intermediate $Q^2$, where GDH sum rules and the
behaviour of $I(Q^2)$ at large $Q^2$ where fullfilled. The model had been
improved in [24].  (Another model with the same goal was
suggested by Soffer and Teryaev [25]).

Since it is known, that at small $Q^2$ the contribution of resonances to
$I(Q^2)$  is of importance, it is convenient to represent $I(Q^2)$  as a sum
of two terms

\be
I(Q^2)  = I^{res}(Q^2) + I^{\prime}(Q^2),
\ee
where $I^{res}(Q^2)$ is the contribution of baryonic resonances.
$I^{res}(Q^2)$ can be calculated from the data on electroproduction of
resonances. Such calculation was done with the account of resonances up to
the mass $W = 1.8 GeV$ [27].

In order to construct the model for nonresonant part $I^{\prime}(Q^2)$
consider the analytical properties of $I(Q^2)$ in $q^2$. As is clear from
(9),(10),  $I(Q^2)$ is the moment of the structure function, i.e. it is a
vertex function  with two legs, corresponding to ingoing and outgoing
photons and one leg with zero momentum. The most convenient way to
study of analytical properties of $I(q^2)$ is to consider a more general
vertex function $I(q^2_1, q^2_2; p^2)$, where the momenta of the photons are
different, and go to the limit $p \to 0, ~q^2_1 \to q^2_2 = q^2$. ~
$I(q^2, q^2_2; p^2)$ can be represented by the double dispersion relation:

$$ I(q^2) = \lim_{q^2_1 \to q^2_2=q^2,p^2\to 0}I(q^2_1,q^2_2;p^2) = \left \{
\int~ds_2 \int~ ds_1 ~\frac{\rho(s_1,s_2;p^2)}{(s_1-q^2_1)(s_2-q^2_2)}
+\right.$$

\be
\left. + P(q^2_1)~\int \frac{\varphi(s, p^2)}{s-q^2_2}ds + P(q^2_2)~\int
\frac{\varphi(s,,p^2)}{s-q^2_1}ds \right \}_{q^2_1=q^2_2=q^2,p\to 0}
\ee
The last two terms in (13)   are the substruction terms in the  double
dispersion relation, $P(q^2)$ is the polynomial. According to (10),
$I(q^2)$ decreases at $\mid q^2\mid \to \infty,$
$P(q^2)=Const$ and the constant subtraction term in (13) is absent. We are
interesting in $I(Q^2)$ dependence in the domain $Q^2 \la 1 GeV^2$. Since
after performed subtraction, the integrals  in (13)  are well converging,
one may assume, that at $Q^2 \la 2 - 3 GeV^2$ the main contribution comes
from vector meson intermediate staties,
so the general form of $I^{\prime}(Q^2)$ is

\be
I^{\prime}(Q^2) = \frac{A}{(Q^2 + \mu^2)^2} + \frac{B}{Q^2 + \mu^2},
\ee
where $A$ and $B$ are constants, $\mu$ is $\rho$  (or $\omega$) mass. The
constants $A$ and $B$  are determined from GDH sum rules at $Q^2=0$  and from
the requirement that at high $Q^2 \gg \mu^2$  takes place the relation

\be
I(Q^2) \approx I^{\prime}(Q^2) \approx \frac{2m^2}{Q^2}\Gamma^{as} (Q^2),
\ee
where $\Gamma^{as}(Q^2)$ is given by (3). $(I^{res}(Q^2)$  fastly decreases
with $Q^2$ and is very small above $Q^2=3 GeV^2$). These conditions are
sufficient to determine in unique way the constant $A$ and $B$ in (14). For
$I^{\prime}(Q^2)$ it follows:

\be
I^{\prime}(Q^2) = 2m^2 \Gamma^{as}(Q^2_m)\Biggl [ \frac{1}{Q^2 + \mu^2} -
\frac{c\mu^2}{(Q^2 + \mu^2)^2}\Biggr ],
\ee

\be
c = 1 + \frac{\mu^2}{2m^2}~\frac{1}{\Gamma^{as}(Q^2_m)}\Biggl [
\frac{1}{4}\kappa^2 + I^{res}(0)\Biggr ],
\ee
where $I_p^{res}(0)=-1.03$, $I_n^{res}(0)=-0.83$ [24].

The model and eq.16  cannot be used at high $Q^2 \ga 5 GeV^2$: one cannot
believe, that at such $Q^2$  the saturation of the dispersion relation (13)
by the lowest vector meson is a good approximation. For this reason there is
no matching of (16)  with QCD sum rule calculations of twist-4 terms.
(Formally, from (16) it would follow $b_{p-n} \approx -0.15,~ b_{p+n} \approx
-0.07$). It is not certain, what value of the matching point $Q^2_m$ should
be chosen in (16). This results in 10\% uncertainty in the theoretical
predictions. Fig.'s 2,3 shows the predictions of the model in
comparison with recent SLAC data [28], obtained at low $Q^2 = 0.5$  and $1.2
GeV^2$  as well as SMC and SLAC data at higher $Q^2$. The chosen parameters
are $\Gamma_p(Q^2_m) = 0.142, ~\Gamma_n(Q^2_m) = -0.061$, corresponding to
$c_p = 0.458, ~c_n = 0.527$  in (16),(17). The agreement with the data,
particularly at low $Q^2$, is very good. The change of the parameters only
weakly influences $\Gamma_{p,n}(Q^2)$ at low $Q^2$.

This work was supported in part by the Russian Foundation of Fundamental
Research Grant 97-02-16131, CRDF Grant RP2-132 and Schweizerische
Nationalfond Grant 7SUPJ048716.

\newpage

\centerline{\large \bf Figure Captions}

\vspace{10mm}
\begin{tabular}{lp{12cm}}
{\bf Fig.~1} & The $Q^2$-dependence of integrals $I_p(Q^2), I_n(Q^2),
I_p(Q^2) - I_n(Q^2)$. The vertical axis is broken at negative values.\\
& \\
{\bf Fig.~2} & The $Q^2$-dependence of $\Gamma_p = \Gamma^{\prime}_p +
\Gamma^{res}_p$ (solid line), described by eqs.(12,16,17). $\Gamma^{res.}_p$
(dotted) and $\Gamma^{\prime}_p$ (dashed) are the resonance and
nonresonance parts. The experimental points are: the dots from E143
[28], the square - from E143 (SLAC) [19], the cross - SMC-SLAC combined data
[1], the triangle from SMC [1].\\
& \\
{\bf Fig.~3} & The same as in Fig.2,but for neutron. The experimental points
are: the dots from E143 (SLAC) measurements on deuteron [28], the square at
$Q^2 = 2 GeV^2$ is the E142(SLAC) [29] data from measurements on polarized
$^3He$, the square at $Q^2 = 3 GeV^2$ is E143(SLAC) [30] deuteron data, the
cross is SMC-SLAC combined data [1], the triangle is SMC deuteron data [1].
\end{tabular}

\begin{figure}
\epsfxsize=10cm
\epsfbox{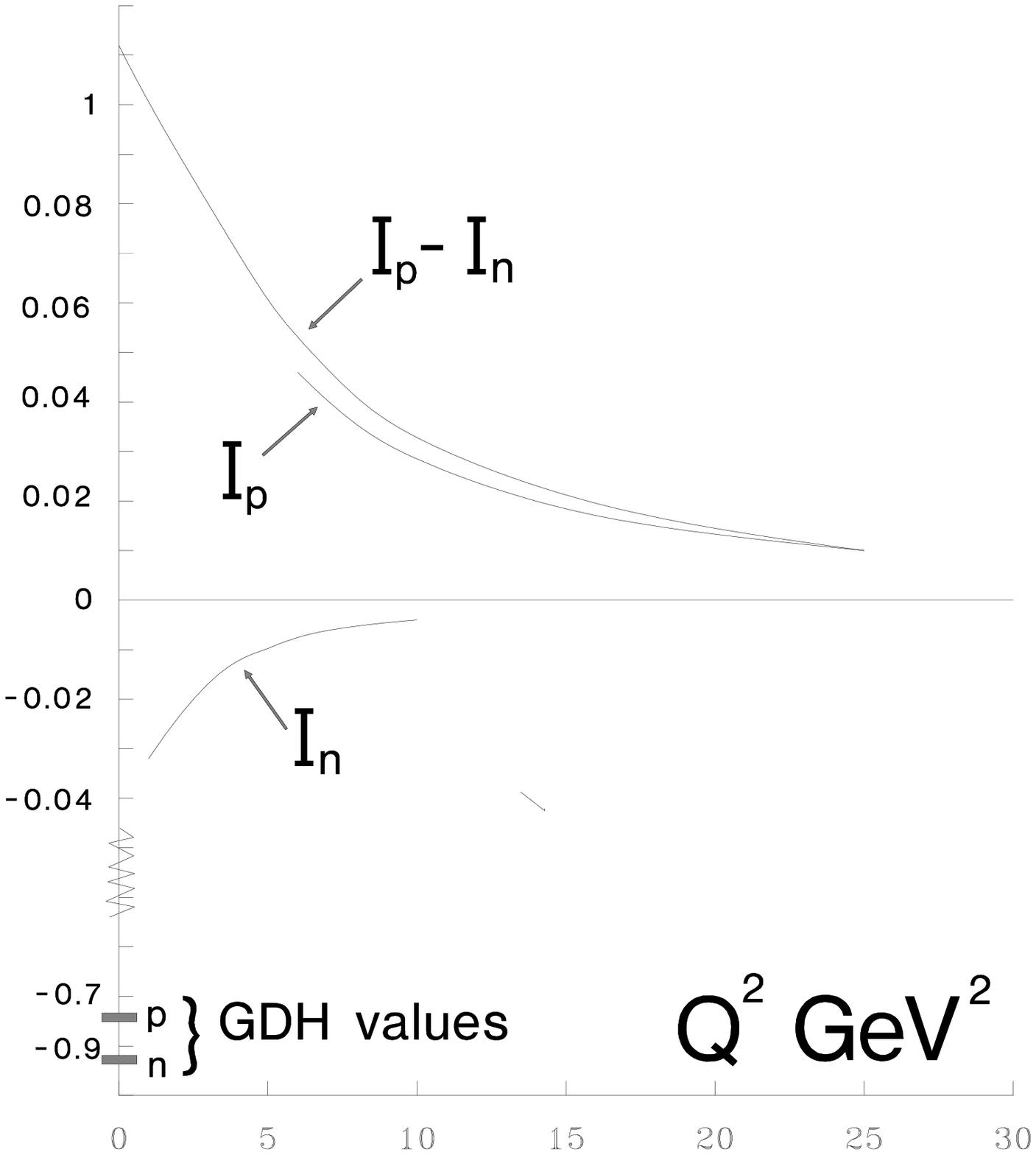}
\caption{}
\end{figure}
\newpage

\begin{figure}
\epsfxsize=10cm
\epsfbox{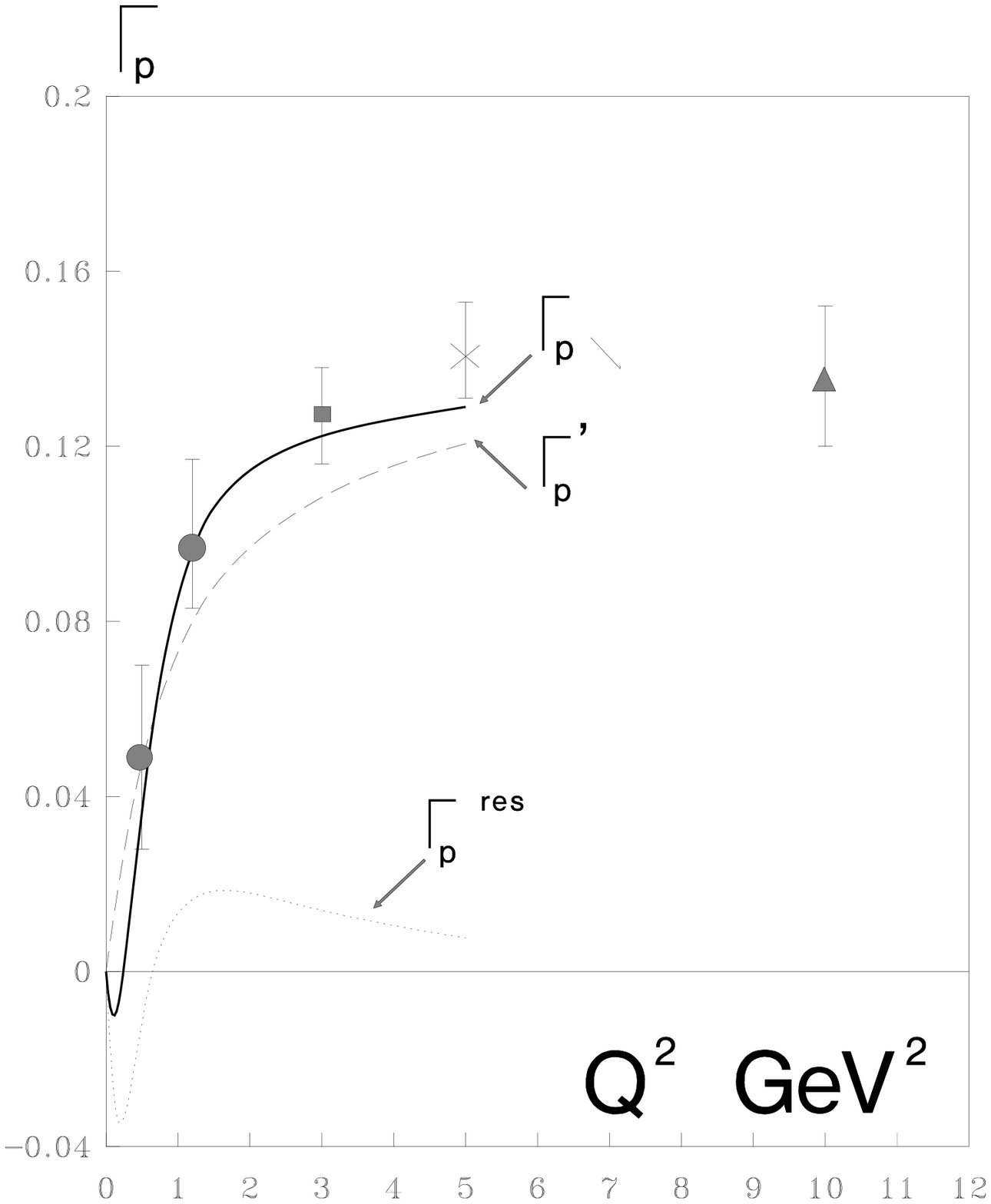}
\caption{}
\end{figure}
\newpage

\begin{figure}
\epsfxsize=10cm
\epsfbox{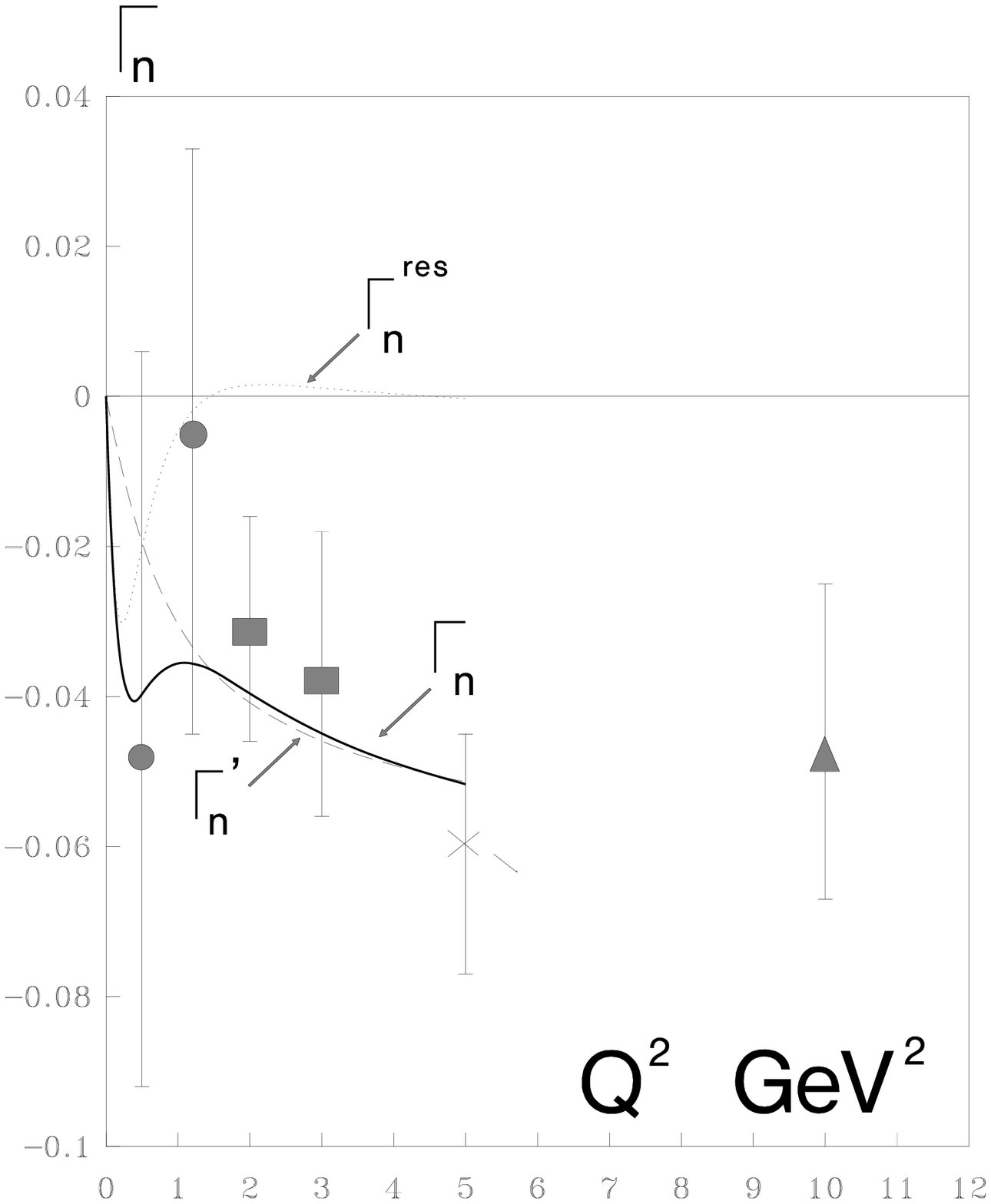}
\caption{}
\end{figure}


\newpage

\begin{thebibliography}{99}
\bibitem{1}SMC-Collaboration, hep-ph/9702005, submitted to Physical
Review D.
\bibitem{2} K.Abe et al (E154 Collab.) Phys.Lett. {\bf B405} (1997) 180.
\bibitem{3} S.Gerasimov, Yad.Fiz. {\bf 2} (1996) 930.
\bibitem{4} S.D.Drell and A.C.Hearn, Phys.Rev.Lett. {\bf 16} (1996) 908.
\bibitem{5} B.L.Ioffe, hep-ph/9704295, Phys.At.Nucl. in press.
\bibitem{6} R.M.Barnett et al., Particle Data Group, Phys.Rev. {\bf D54} (1996) 1.
\bibitem{7} J.Kadaira et al., Phys.Rev. {\bf D20} (1979) 627;
Nucl.Phys. {\bf B159} (1979) 99, {\bf 165} (1980) 129.
\bibitem{8} S.A.Larin and J.A.M.Vermaseren, Phys.Lett. {\bf 259} (1991) 345.
\bibitem{9} S.A.Larin, Phys.Lett. {\bf 334} (1994) 192.
\bibitem{10} S.A.Larin, T.van Ritbergen and J.A.M.Vermaseren,
preprint NIKHEF-97-011 (1997).
\bibitem{11} A.L.Kataev, Phys.Rev. {\bf D50} (1994) 5469.
\bibitem{12} S.Y.Hsueh et al., Phys.Rev. {\bf D38} (1988) 2056.
\bibitem{13} R.D.Carlitz, J.C.Collins and A.H.Mueller, Phys.Lett. {\bf B
214} (1988) 229.
\bibitem{14} S.D.Bass, B.L.Ioffe, N.N.Nikolaev and A.W.Thomas,
J.Moscow Phys.Soc. {\bf 1}
(1991) 317.
\bibitem{15} I.I.Balitsky, V.M.Braun and A.V.Kolesnichenko,
Phys.Lett. {\bf 242} (1990) 245,

Errata {\bf B318} (1993) 648.
\bibitem{16} A.Oganesian, hep-ph/9704435, Phys.At.Nucl., in press.
\bibitem{17} M.J.Alguard et al, Phys.Rev.Lett. {\bf 37} (1976) 1261;
{\bf 41} (1978) 70. G.Baum, ibid {\bf 51} (1983) 1135.
\bibitem{18} J.Ashman et al, Phys.Lett {\bf 206} (1988) 364; Nucl.Phys. {\bf
B328} (1989) 1.
\bibitem{19} K.Abe et al. (SLAC E143 Collaboration),
Phys.Rev.Lett. {\bf 74} (1995) 346.
\bibitem{20} D.Adams et al, Phys.Lett. {\bf B329} (1994) 399.
\bibitem{21} I.Balitsky, X.Ji, Phys.Rev.Lett. {\bf 79} (1997) 1225.
\bibitem{22} A.Saalfeld, G.Piller, L.Mankiewicz,
Preprint TUM/T39-97-19, hep-ph/9708378.
\bibitem{23} M.Anselmino, B.L.Ioffe and E.Leader, Sov.J.Nucl.Phys. {\bf 49}
(1989) 136.
\bibitem{24} V.Burkert and B.L.Ioffe, JETP, {\bf 105} (1994) 1153.
\bibitem{25} J.Soffer and O.Teryaev, Phys.Rev. {\bf D51} (1995) 25.
\bibitem{26} B.L.Ioffe, V.A.Khoze, L.N.Lipatov, Hard Processes, v.1,
North Holland, Amsterdam, 1984.
\bibitem{27} V.Burkert and Z.Li, Phys.Rev. {\bf 47} (1993) 46.
\bibitem{28} K.Abe et al., (SLAC E143 Collaboration) Phys.Rev.Lett.
{\bf 78} (1997) 815.
\bibitem{29} P.L.Anthony et al. (SLAC E142 Collaboration),

Phys.Rev. {\bf D54} (1996) 6620.
\bibitem{30} K.Abe et al. (SLAC E143 Collaboration),
Phys.Rev.Lett. {\bf 75} (1995) 25.
\end{thebibliography}
\end{document}